# A pipeline and comparative study of 12 machine learning models for text classification

Annalisa Occhipinti [a,b,d+], Louis Rogers [a+], Claudio Angione [a,b,c,d*]

[a] School of Computing, Engineering and Digital Technologies, Teesside University, Middlesbrough, TS1 3BA, UK.
[b] Centre for Digital Innovation, Teesside University, Middlesbrough, TS1 3BA, UK.
[c] Healthcare Innovation Centre, Campus Heart, Teesside University, Middlesbrough, TS1 3BX, UK.
[d] National Horizon Centre, Teesside University, Darlington, DL1 1HG, UK

[+] Equal contribution

Email addresses of all authors:
AO: a.occhipinti@tees.ac.uk
LR: louismb@msn.com
CA: c.angione@tees.ac.uk

* Corresponding author: Claudio Angione

School of Computing, Engineering and Digital Technologies
Teesside University
Borough Road, Middlesbrough, North Yorkshire
TS1 3BA
UK

Email address: c.angione@tees.ac.uk
Telephone: +44 1642 342659

# Abstract

Text-based communication is highly favoured as a communication method, especially in business environments. As a result, it is often abused by sending malicious messages, e.g., spam emails, to deceive users into relaying personal information, including online accounts credentials or banking details. For this reason, many machine learning methods for text classification have been proposed and incorporated into the services of most email providers. However, optimising text classification algorithms and finding the right tradeoff on their aggressiveness is still a major research problem.

We present an updated survey of 12 machine learning text classifiers applied to a public spam corpus. A new pipeline is proposed to optimise hyperparameter selection and improve the models' performance by applying specific methods (based on natural language processing) in the preprocessing stage.

Our study aims to provide a new methodology to investigate and optimise the effect of different feature sizes and hyperparameters in machine learning classifiers that are widely used in text classification problems. The classifiers are tested and evaluated on different metrics including F-score (accuracy), precision, recall, and run time. By analysing all these aspects, we show how the proposed pipeline can be used to achieve a good accuracy towards spam filtering on the Enron dataset, a widely used public email corpus. Statistical tests and explainability techniques are applied to provide a robust analysis of the proposed pipeline and interpret the classification outcomes of the 12 machine learning models, also identifying words that drive the classification results. Our analysis shows that it is possible to identify an effective machine learning model to classify the Enron dataset with an F-score of 94%. All data, models, and code used in this work are available on GitHub at https://github.com/Angione-Lab/12-machine-learning-models-for-text-classification.

# Introduction

While internet users across the world are accepting a socially connected world, the importance of managing unwanted, and potentially damaging text-based campaigns (such as spam) requires text classification methods to be constantly optimised in order to protect the users.

Online platforms, known as social media, are another form of communication derived from electronic email. For such platforms, spammers have developed sophisticated campaigns that lead unsuspecting users into falling and providing personal details that could relate to banking and online credentials. Recent popular methods developed from spam, known as phishing or malware delivery emails, have been seen on the rise over the past years due to their low risk and high rewards outcomes. This is due to the design of these campaigns and the implementation of obfuscation techniques such as Botnets (Bertino & Islam, 2017), used to carry out the campaign without the need for human monitoring. Therefore, this activity is widely used as a means of producing money and it is seen as illegal in all terms by the law.

Today, more than 86% of messages and emails received by users are considered spam, specifically a collection of phishing, malware delivery and spam emails (Bhardwaj *et al*., 2020). Thus, spam is still a major problem, not just towards the public, but also towards organisations. Indeed, any breach of internal accounts could present damages towards their image as well as governing bodies presenting fines and sanctions as a repercussion.
Due to the increasing complexity of spam campaigns, machine learning, natural language processing (NLP), and deep learning techniques need to be further developed and carefully optimised to minimise the incidence of the spam problem.

Before applying any machine learning methods to classify the data into spam or ham (non-malicious email), a preprocessing stage is required. This will produce a data structure for feature extraction in order to reduce the noise within the data. Techniques such as stop word and hypertext mark-up language tag removal (HTML) are commonly used to remove words that have no value, thus reducing the possibility of false classification. This is known as the preprocessing stage during text classification (Méndez, 2005).

NLP methods such as stemming and lemmatisation can reduce the noise and dimensionality of the feature extraction and vectorisation of the initial data. The method of stemming removes suffixes and prefixes applied to words; an algorithm known as porter stemmer (Porter, 1980) is a favoured technique for removing suffixes and has been used for information retrieval (Alotaibi and Gupta, 2018). The method of lemmatisation attempts to reduce a word to its simplest form e.g., 'impacted' to 'impact'. This method can be used as a preprocessing method for reducing noise and producing a reliable structure for feature extraction.

Over many years, researchers have applied machine learning for text classification problems. The favoured and most commonly deployed method is Naïve Bayes (NB), as it has been reported to be the best performing classifier, in terms of running time and accuracy (Metsis *et al*., 2006a; Fu *et al., 2021*). More recently, other algorithms like decision trees, support vector machines (SVM), nearest neighbours, and neural networks have also been reported to be effective in the text classification domain (Trivedi, 2016).

The purpose of this work is to test 12 machine learning classifiers and identify whether the classification outcome could be improved based on the dataset size, whilst exploring and attempting to optimise the hyperparameters, specifically with neural networks, nearest neighbours, and SVM algorithms. The performance of each classifier has been measured in terms of precision, recall, and F-score whilst recording the time required for each algorithm to process the classification task.

The outline of this paper is as follows. The next section reports a discussion on research related to the text classification domain, specifically relating to spam emails, exploring the available tools and outlining why our proposed pipeline can help address areas that are reported as concerns. The Methods section covers the methodology, the stages of preprocessing, feature extraction, vectorisation and classification. It also provides the mathematical model behind each algorithm, and how its implementation was achieved in Python. The Results section reports the outcomes of the investigation, reviewing and discussing the outcomes. This section also includes the statistical analysis carried out to assess the performance of the 12 machine learning models and interpret the classification outcomes. Finally, the Conclusion section summarises the investigation, stating any concerns, and proposing areas to consider and take forward as future research issues in the text classification research domain.

## 1.1 Related work

The most common machine learning algorithm based on text classification and spam filters deployed and released under open-sourced license is Naïve Bayes (NB), originally derived from Bayes' theorem. Metsis *et al.* (2006a) investigated and discussed the most appropriate NB classifier by using a public corpus known as Enron to examine the performance of derived NB algorithms including multi-variate Bernoulli NB, multinomial NB, and flexible Bayes. The outcome found multinomial NB to be the best performing classifier. A similar study was performed by Jia *et al*. (2012), who applied three classification methods (rule-based technique, decision tree algorithm, and SVM) to classify spam websites that were found in search engine queries. The authors reported that rule-based and decision tree applications were not effective since they were unable to create links between multiple web pages. However, SVM was found to outperform the other two methods in classifying spam and normal websites, while rule-based and decision tree showed similar performance.

A commonly used algorithm for email classification is neural networks (NN). One of the first applications of NN for email classification included the use of 'a bag of words' technique to filter and produce usable features (Clark *et al.,* 2003). Specifically, a variant of NN called multi-layer perceptron was implemented, using the backpropagation algorithm to train the model and determine the classification outcome. The proposed model was run on a public spam corpus known as Ling-spam (Androutsopoulos, 2003) and it was reported to outperform several other algorithms.

NN models have been integrated with other models to improve the classification performance. The approach presented by Manjusha *et al.* (2010) used a combination of NB and NN to process both header and body information from an email in order to classify the outcome as spam or ham. The outcome found the combined classifier to perform well, with an average F-score of 98%. However, concerns about the dataset used in this research have been raised. Indeed, the data was collected from the personal inbox of the author, which could indicate a bias towards the data.

Several machine learning and deep learning classifiers have been applied on the Enron public corpus (Trivedi, 2016; Metsis, 2006b), including Bayesian, NB, SVM, and a Java variant of decision trees known as J48 (Patil, 2013). While it is common to see these algorithms applied to a classification problem, Trivedi (2013) proposed a boosting method known as AdaBoost or bootstrapping**.** The results showed a small increase in accuracy for Bayesian and NB. However, the time required was vastly increased resulting in the SVM method having the best outcome in terms of classification performance and low false-positive rate.

A weighting method was proposed by George *et al.* (2015) during the feature extraction process, called term frequency-inverse document frequency (TF-IDF). The method was applied to reduce the dimensionality of the feature space and the noise contained within the dataset, therefore drastically improving the outcome of the classifier. Multinomial NB (MNB) and SVM classifiers were applied with ten-fold cross-validation during the classification task. The results showed SVM to perform better than MNB in terms of accuracy. However, SVM took longer to process the data than MNB due to the complex mathematics associated with the algorithm.

Fette *et al.* (2007) presented the comparison of various classifiers associated with random forest, SVM, Bayesian, and rule-based approaches for detecting phishing emails. An issue with phishing compared to spam is due to the design of phishing emails. The method of phishing consists of leading a user into thinking that an email is legitimate from a company. As expected, this can create difficulty in misclassification between legitimate and counterfeit emails, increasing the chance of false negatives. However, the authors incorporated the header and URLs contained within each email and they processed them accordingly to address this issue. A possible solution was found by using text classification while implementing specialised filters to generate usable features. The proposed system, known as 'PILFER', was effective with the dataset used, but it would require further testing and validation if made available for general users. To address this issue, Feroz *et al.* (2014) proposed a method that uses various machine learning classifiers to detect phishing activity by inspecting extracted features related to a URL. During the feature extraction and selection stage, Chi-squared test and information gain were used to process and select the most useful features for classification. The following algorithms were assessed in the research work: logistic regression, J48 SVM, Random Forest, NB, and BayesNet. The outcome found logistic regression to be the most accurate over the three size tests using k-fold validation in the ratio 1:1, 4:1, and 10:1. However, the false-positive rate for logistic regression was significantly higher than for the other algorithms.

One of the most recent applications for text classification is based on the evaluation of machine learning classification methods for detecting streamed twitter spam using the developer tools provided by Twitter. Chen *et al.* (2015) collected over 600 million tweets over a period of three weeks. While spam classifiers associated with social media platforms have unique issues to identify, the text processing method is similar in terms of email/spam, using machine learning to classify whether an email or tweet is spam or ham. The results showed deviation in the classification outcomes based on the data collected from various days; this is known as the streaming problem associated with spam tweet campaigns and it does not relate with email classification. However, classifiers such as Bayes Net, J48, SVM, and kNN were considered the best performing algorithms. Mujtaba *et al.* (2017) provided a critical review of email classifier algorithms proposed between 2006 and 2016. The report was divided into specific sections such as email classification area, applied datasets, and classifiers used. The most common email classification area was "applied supervised learning approach", with forty papers. Enron, SpamAssassins, and TERC spam corpora were found to be the most common datasets, while multi-folder classification methods favoured PU, Phishing Corpus, and

Enron. The most frequently used classifying techniques deployed were SVM, decision trees, and NB. Finally, the most commonly used performance measures were precision, recall, accuracy, and F-score, closely followed by false-positive rate, false-negative rate, and error rate.

Machine learning and NLP have also been applied to extract composite email features from the Enron dataset (George and Vinod, 2018). Dimensionality algorithms were used to rank the extracted features (including character-based, word-based, tag-based, and structured based features) before applying the machine learning algorithms. Experiments showed that SVM was the best performing algorithm out of the 5 algorithms tested, which only included Multinomial Naïve Bayes and SVM (with four different kernels). More recently, machine learning algorithms have been merged with evolutionary and adaptable spam models for spam classification tasks. Specifically, Faris *et al.* (2019) developed an intelligent decision system based on genetic algorithms and random weight networks able to automatically identify the most relevant features of spam emails. However, the work does not include any preprocessing step or feature selection analysis.

Moreover, only a few studies have focussed on identifying the influence of the features and interpreting the machine learning results. Therefore, the aim of this paper is three-fold: (i) to investigate the impact of different feature sized datasets and hyperparameters on the performance of 12 machine learning classifiers, (ii) to statistically analyse and explain the classification outcomes, and (iii) to provide a Python script that can be easily adapted to different spam datasets and scenarios.

## Materials & Methods

In this paper, we present a classification pipeline that compares 12 supervised machine learning classifiers. We aim to explore noise reduction tasks like lemmatisation, stop words, and HMTL tag removal. We also evaluate hyperparameter optimisation and variants of machine learning classifiers that could impact the classification outcomes using the well documented Enron corpus (Metsis, 2006b), which will be further discussed in the next section. Figure 1 shows the pipeline proposed in this paper. (a) First, the dataset is selected and (b) train and test sets are allocated (70% train and 30% test). (c) Then, in order to reduce the noise in the data, a preprocessing stage is applied. This includes lemmatisation, stop words, and HTML tags removal. (d) Features are then extracted for generating the dictionary and the matrices for the train and test sets. (e) These matrices are used by the 12 machine learning algorithms to fit the data and predict the classification outcomes. (f) Finally, statistical models are applied to assess the significance of the classification results and provide an interpretation of the classification outcomes. The Python script to run the models is available on GitHub at https://github.com/Angione-Lab/12-machine-learning-models-for-text-classification.

## 2.1 Enron Dataset

The dataset used in this study is the Enron spam corpus (Metsis, 2006b). Table 1 outlines the total number of ham and spam emails in each subset of the corpus. To avoid bias in the dataset, the proportion has been set to 0.5, allowing an equal selection of ham and spam emails. However, ingesting the total dataset would nullify this request due to more spam emails available (17171 spam emails compared to 16545 ham emails). Therefore, we selected an equal number of spam/ham emails by randomly extracting 16545 spam emails, allowing an equal proportion of the two classes. The dataset was then split into train and test sets (70% and 30%, respectively, Figure 1(a)). For this investigation, we used the preprocessed data rather than the raw data; however, noise items like HMTL tags and stop words were still present.

## 2.2 Preprocessing and feature selection

The purpose of the preprocessing steps is to reduce the noise contained in the data and improve the selection of features to be used for the classification task (Figure 1(c)). The proposed preprocessing steps apply the removal of stop words, HTML tags, and single letters or numbers. We applied an NLP method known as lemmatisation (Cambria, 2014) that replaces inflected words with their simplest form, e.g., changing 'following' into its simplest form 'follow'. All the words that could be classified as noise (since common to both email and spam instances) were removed. These include words like *subject*, *cc*, and *to*. The word *Enron* was also removed to provide more general and less company-specific results.

The matrix generation phase took place once the dataset had been processed (Figure 1(d)). Then, the most occurring features were selected as a dictionary, which was used to translate the features into a matrix using vectors. The process of recording vectors is similar to Word2Vec (Mikolov, 2013). Iterations take place between a generated matrix containing the features labelled as *x*, and the dataset (either train or test sets) labelled as *y*. For each word found in dataset *y*, the model adds a 1 to label *x*, thus building a matrix calculation of each email.

## 2.3 Machine learning algorithms

Using the set of extracted features, machine learning algorithms were applied to fit the data and generate the classification outcomes. Specifically, 12 machine learning algorithms were deployed in our work: naïve Bayes (multinomial, Gaussian, and Bernoulli), supporting vector machine (linear, polynomial, sigmoidal, and radial basis kernels), nearest neighbours, multinomial perceptron neural network, logistic regression, random forest, and extreme gradient boost. A list of the parameters used for each method is reported in Table 2 and discussed in the following sections.

*Naïve Bayes*

Naïve Bayes (NB) algorithms are supervised learning models that apply Bayes' theorem with the "naïve" assumption of independence between every pair of features (Mitchell, 1999). We analyse below three different NB algorithms included in the paper: Multinomial NB, Gaussian NB, and Bernoulli NB.

*Multinomial Naïve Bayes*

Let the set of classes be denoted by $C$. Let $N$ be the size of the dictionary. Multinomial NB (MNB) classifier assigns a test document $t_i$ to the class that has the highest probability $Pr(c|t_i)$ given by

$$Pr(c|t_i) = \frac{\Pr(c)Pr(t_i|c)}{\Pr(t_i)}. \qquad c \in C \qquad (1)$$

$\Pr(c)$ is the class prior, and it is equal to the number of documents belonging to class $c$ ($N_c$) divided by the total number of documents in the dataset ($N$), $Pr(c) = \frac{N_c}{N}$. $\Pr(t_i)$ is the probability of the input instance, which is independent of the classes. $Pr(t_i|c)$ is the probability of observing the document $t_i$ in a given class $c$, and it is calculated as

$$Pr(t_i|c) = \left(\sum_n f_{ni}\right)! \prod_n \frac{Pr(w_n|c)^{f_{ni}}}{f_{ni}!}, \qquad (2)$$

where $f_{ni}$ is the number of words $n$ in the document $t_i$, and $Pr(w_n|c)$ is the probability of word $n$ given class $c$. This is estimated as

$$\widehat{Pr}(w_n|c) = \frac{F_{nc} + 1}{N + \sum_{x=1}^{N} F_{xc}}, \qquad (3)$$

where $F_{xc}$ is the number of words $x$ in the training dataset belonging to class $c$, and $N$ is the size of the dictionary. The term +1 at the numerator is used to avoid the zero-frequency problem (McCallum, 1998).

The multinomial condition captures the frequency information of each word to improve the classification outcomes. Maximum a posteriori estimation (Rish, 2001) is commonly used to estimate the parameters in the NB model, including $\Pr(c)$ and $Pr(t_i|c)$. MNB has been widely used in text classification problems resulting in one of the most effective algorithms in spam detection (Juan, 2002; Lewis, 1998; Panda, 2010).

*Gaussian Naïve Bayes*

Gaussian NB implements the classification algorithm by defining the likelihood of observing instance $t_i$, given class $c$, as

$$Pr(t_i|c) = \frac{1}{\sqrt{2\pi\sigma_y^2}} \exp\left(-\frac{(t_i - \mu_y)^2}{2\pi\sigma_y^2}\right), \quad (4)$$

where the parameters $\sigma_y$ and $\mu_y$ are estimated by maximum likelihood. Because of its simplicity and being extremely fast, Gaussian NB has been applied in several prediction problems (Cao, 2003; Murakami, 2010; Raizada, 2013).

*Bernoulli Naïve Bayes*

In the Bernoulli NB model, features are considered as independent binary variables (Booleans) describing inputs. If $x_i$ is the Boolean variable expressing the occurrence or absence of the $i-th$ word from the dictionary, then the likelihood of observing document $x$, given a class $c$, is defined as follows

$$Pr(\boldsymbol{x}|c) = \prod_{i=1}^{n} p_c^{x_i} (1 - p_c)^{(1-x_i)}, \quad (5)$$

where $p_c^{x_i}$ is the probability of observing the term $x_i$ in the class $c$, and $n$ is the number of words in the dictionary.
This model has been widely used for classifying short texts since it has the benefit of modelling the absence of terms (McCallum, 1998).

*Support Vector Machine*

SVM is a useful model for pattern classification. The method is based on a statistical theory proposed by Vapnik (2013), and it can be applied to both linearly separable features and non-linearly separable features.

Given the training set $(x_n, y_n), n = 1, ..., N$, where $x_n$ is a vector containing the features associated with each instance $n$, and $y_n$ is the class label for each instance $n$, the SVM classifier defines the "maximum-margin hyperplane" separating the classes. The hyperplane is defined so that the distance between the hyperplane and the nearest point $x_n$ from either group is maximised. In classification problems, specifically regarding spam classification, classes are set as labels +1 and -1 retrospectively to spam and ham outcomes.
We present below the details of the SVM models applied in our work. The main difference between the models consists in how each model creates the hyperplane (decision boundary between the classes) based on the mathematical definition of the kernel function.

*SVM - Linear kernel*

The linear SVM classifier uses a hypothesis vector $w$, normal to the hyperplane, and a bias $b$ to classify an instance $x$ by using a prediction class label $f(x)$ defined as

$$f(x) = sign(\langle w, x \rangle + b). \tag{6}$$

$w$ defines the separating hyperplane by minimising the objective function:

$$\tau(w, \xi) = \frac{1}{2}\|w\|^2 + c \sum_{i=1}^{n} \xi_i \tag{7}$$

under the constraints that $\forall i = \{1 ... n\} : y_i(\langle w, x \rangle + b) \geq 1 - \xi_i \geq 0$.
The objective function describes each slack variable $\xi_i$ to show the recorded error that the classifier makes on the instance $x_i$. Minimising the sum of the slack variables $\xi_i$ corresponds to minimising the loss function on the training data. Minimising the term $\frac{1}{2}\|w\|^2$ corresponds to maximising the margins between the two classes. These optimization tasks are in conflict and the parameter $c$ is used as a trade-off to determine the importance to give to each task (Scully, 2007).

Previous research (Almedia, 2011; Ott, 2011; Scully, 2007; Svore, 2007) into text classification that has two distinct or solvable problems shows linear SVM to be an effective classifier. The linear SVM classifier used in our work is based on the classification method described in (Scully, 2007) and (Joachims, 1998).

*SVM - Polynomial kernel*

Polynomial kernel SVM is based on a similar approach as the linear kernel, but it does not rely only on one given feature of the input samples to determine their similarity, but also on combinations of these. For a degree-$d$ polynomial, the kernel function is defined as

$$K(\mathbf{u}, \mathbf{v}) = (\mathbf{u} \cdot \mathbf{v} + 1)^d, \qquad (8)$$

where $\mathbf{u}$ and $\mathbf{v}$ are vectors in the input space, i.e., feature vectors computed from train or test samples. The decision function is defined as

$$f(x) = \sum_{i=1}^{n} y_i \, \alpha_i K(\mathbf{x}, \mathbf{x}_i), \qquad (9)$$

where $\mathbf{x}_i$ is the image of a support vector in the input space, and $\alpha_i$ is a parameter describing the weight of a support vector in the feature space.
The most common degree for polynomial SVM is $d = 3$, since larger degrees tend to overfit the data (Cortes, 1995). Hence, in our test, we set $d = 3$.

*SVM - Sigmoidal kernel*

The sigmoidal kernel SVM is a popular method because of its origin from neural networks. In fact, its kernel is equivalent to a two-layer perceptron neural network. In the sigmoidal SVM method, the kernel is defined as

$$K(\mathbf{u}, \mathbf{v}) = \tanh(\mathbf{u} \cdot \mathbf{v} + r), \qquad (10)$$

where $r$ is a kernel parameter. Even if the model is only defined when the kernel matrix is positive semi-definite (PSD), it has been found that the model performs well even when the matrix is not PSD (Lin, 2003; Boughorbel, 2005).

*SVM - Radial basis function kernel*

The radial basis function (RBF) kernel has been commonly used in SVM classification problems (Chang, 2010; Vert, 2004). The RBF kernel on two samples $\boldsymbol{u}$ and $\boldsymbol{v}$, represented as feature vectors in the input space, is defined as

$$K(\boldsymbol{u},\boldsymbol{v}) = exp\left(-\frac{\|\boldsymbol{u}-\boldsymbol{v}\|^2}{2\sigma^2}\right), \qquad (11)$$

where $\|\boldsymbol{u}-\boldsymbol{v}\|^2$ is the squared Euclidean distance between the feature vectors, and $\sigma$ is a regularisation parameter. The value of the RBF kernel decreases with the distance and ranges between zero and one, and the RBF kernel maps the data into an infinite-dimensional space. As $\sigma^2 \to \infty$, SVM with the RBF kernel and penalty parameter $c$ approaches linear SVM with the penalty parameter $c/(2\sigma^2)$. This result implies that with a suitable parameter selection, the accuracy of the RBF kernel can be at least as good as the linear kernel (Keerthi, 2003).

*k-Nearest Neighbours*

Nearest Neighbours is one of the most commonly applied classifiers for text classification problems including spam detection (Aci, 2010; Cunningham,2007; Soonthornphisaj, 2002). The specific variant known as *k-Nearest Neighbours* (kNN) queries the closest *k* neighbours and determines the outcome of a class based on the neighbouring items that are associated with a class. We explored three algorithms associated with kNN: brute force, ball tree, and k-dimensional (KD)-tree. These apply different methods for measuring the distance between items and, ultimately, to determine the classification outcomes. In particular, brute force is based on the simple intuition of calculating the Euclidean distance between the instance to be classified and all the instances in the training set. Then the point is added to the class of the majority points in the *k*-nearest reference elements. Ball tree assumes that the data is in a multidimensional space and creates the nested hyperspheres to classify each point based on the *k*-nearest reference elements in each hypersphere. KD-tree has been developed to improve the running time of the algorithm. It is based on the idea that if an instance P is distant from Q, and Q is very close to R, then P and R must be distant (without calculating their exact distance). Even if this might not be always a correct assumption, KD-tree has been widely used to reduce the computational cost of kNN (Muja, 2009).

Euclidean distance is commonly used to calculate the distance between the neighbouring classes, where the majority of votes from all neighbouring instances dictates the outcome of the item in question. The Euclidean distance between two feature vectors $A = (x_1, x_2,..., x_k)$ and $B = (x_1, x_2,..., x_k)$ is defined as follows

$$dist(A, B) = \sqrt{\sum_{i=1}^{k}(x_i - y_i)^2}. \tag{12}$$

The final class is assigned based on the distance from *k* neighbours. If *k*=1, the item is assigned to the same class as the closest neighbour.

Table 2 shows the values used for the hyperparameters in the kNN classifier. Other than testing the effects of the three different algorithms (brute force, ball tree, and KD-tree), we also set the parameter *k* equal to its default values, i.e., *k* = 5 neighbours. *k* is used to determine the class of the new point based on the class of the closest *k* neighbours. Leaf size is a parameter passed to ball tree and KD-tree that affects the construction, query and memory required to store the tree. This parameter controls the number of samples at which a query switches to brute-force. This allows both algorithms to approach the efficiency of a brute-force classifier. We set this parameter equal to 10. Finally, we set the power parameter *p* equal to 1. This parameter regulates the Minkowski metric. Specifically, when *p=1* the Manhattan distance is used ($L_1$-norm).

*Multilayer Perceptron Neural Network*

Multilayer Perceptron Neural Network (MPNN) is an effective method for problems that can be defined either linearly or non-linearly based on two classes (for example, ham and spam, Gardner, 1998). MPNN consists of a system of interconnected nodes (neurons) that represent a non-linear mapping between an input vector and an output vector. Weights and output signals are used to connect the nodes. These signals are functions of the sum of the inputs to the node modified by a non-linear activation function. Through the superimposition of many non-linear functions, the MPNN approximates extremely non-linear functions. The architecture of an MPNN generally consists of several layers of neurons, input and output layers, and the hidden layers in between.

The objective of the training process for the MPNN is to find an unknown function *Y= f(X)*, where *X* is a matrix of size [*n,k*], *Y* is a matrix of size [*n,j*], *n* is the number of training inputs, *k* is the number of input nodes in the network, and *j* is the number of output nodes (or classes). During the training phase, the function *f(X)* is optimised such that its output is as close as possible to the target value *Y*.

MPNN is a non-linear feed-forward network, and previous research (Alsmani, 2009; Goh, 2013; Soranamageswari, 2010) has shown that MPNN is capable of handling difficult and intensive problems such as spam image detection and spam web content. The specific variant used in our work is called Limited-memory Broyden Fletcher-Goldfarb-Shanno (LBFGS) algorithm. This method is an approximation of Newton's method and improves the training efficiency of back-propagation neural network algorithms by adaptively modifying the initial search direction. The reader is referred to Nawi *et al.* (2006) for a detailed explanation of the MPNN algorithm that applies LBFGS. In terms of parameters regulation, we trained the models tuning the number of neurons, the maximum number of iterations before reporting the classification outcomes, and the number of hidden layers (Table 2). The effects of these parameters in terms of performance are reported in the Results section.

*Logistic Regression*

Logistic Regression (LR) has been considered effective at binary classification problems by creating a linear (or sigmoid line) to define two classes based on the dataset provided. Previous research has shown that LR performs well in spam detection problems (Chang, 2008; Jindal, 2007; Genkin, 2012; Ravikumar, 2007). One of the advantages of logistic regression is that it can be easier to interpret compared to other machine learning algorithms. When the dataset has a large number of variables, regularisation methods can be used as a feature selection tool, to further facilitate the interpretation of the results (Lipton, 2018; Magazzu, 2021). In LR, given $N$ instances $x_i$, $i = 1, \ldots, N$, the $m$ features of the input instance $x_i = (x_{i1}, x_{i2}, \ldots, x_{im})$, are combined linearly using coefficients $\beta_0$ and $\boldsymbol{\beta} = (\beta_{i1}, \ldots, \beta_{im})$ to predict the classification outcome $y_i$. Specifically, given an input instance $x_i$, the probability that $y_i$=1 is denoted by $p(x_i)$ and is modelled with the standard logistic regression model as follows

$$p(x_i) = \frac{e^{(\beta_0 + \beta * x_i)}}{1 + e^{(\beta_0 + \beta * x_i)}}. \tag{13}$$

The aim of LR is to find the best parameters $\beta_0$ and $\boldsymbol{\beta}$ to determine the best fitting model that describes the relationship between the input features and the predicted class value. In our test, we used LR with the LBFGS solver, which is a variant of Newton's method as explained in the previous section. For a comprehensive review of the logistic regression algorithm, the reader is referred to Genkin *et al.* (2012).

*Random Forest*

Random Forest (RF) has been widely used for classification problems (Boulesteix, 2011; Touw, 2012). This is due to the high prediction performance of the model, which also provides information on the variables' importance for classification purposes. The algorithm is based on a large number of decision trees where each tree generates a classification outcome and the forest gives the final classification outcome based on the number of votes (over all the trees in the forest). The most commonly used RF algorithms are based on the so-called decrease of Gini impurity for split selection (Zhang, 2017). Gini impurity measures how often a randomly chosen element from the set would be incorrectly labelled if it was randomly labelled according to the distribution of labels in the subset. Hence, Gini impurity is based on the probability of a new item to be incorrectly classified at a specific node in the decision tree, based on the training data. Table 2 shows the parameters set for our test, i.e., the number of trees in the forest, which determines the number of trees the algorithm will build before taking the maximum voting or averaging the predictions, and the minimum number of samples to split an internal node, which provides a lower bound for the number of samples required for a split.

*Extreme Gradient Boost*

Extreme Gradient Boost (XGBoost) is an algorithm based on a parallel tree-boosting system designed for optimising both speed and performance. It is built on the principle of the gradient boosting framework for classification and regression (Friedman, 2001). XGBoost combines weak "learners" into a stronger "learner" using iteration. At each stage $m$ of the gradient boosting iteration, we might assume that there is some imperfect model $F_m$. XGBoost improves $F_m$ by constructing a new model that uses the residual of $F_m$ to construct a new model $F_{m+1}$. This new model attempts to correct any errors of the previous one. XGBoost is a gradient descent algorithm that supports various objective functions, including regression, classification, and ranking. The algorithm is optimised for sparse input for both tree booster and linear booster, and it has shown better performance than other machine learning methods in different applications (Chen, 2016).

**2.4 Hyperparameter selection and performance metrics**

Table 2 outlines the final parameters used for our study. Specifically, the hyperparameters reviewed in this paper affect support vector machine (SVM), *k*-nearest neighbours (kNN), multilayer perceptron neural network (MPNN), logistic regression (LR), and random forest (RF). In our pipeline, we included the following kernels for SVM: linear, polynomial (degree 3), sigmoidal, and radial basis. The number of maximum iterations was set equal to 5, and the penalty parameter equal to 1. For *k*-nearest neighbours, the best distance algorithm (brute force, ball tree, and KD-tree) was identified and selected for the final tests, as discussed in the Results section. Weight type, power parameter, leaf size, and neighbours count were set equal to 10, 1, and 5, respectively. The hyperparameters tuned for MPNN included the number of hidden layers, number of neurons, and number of maximum iterations. A discussion on how the optimal parameters were set is provided in the Results section. The parameters set for LR were the solving algorithm (LBFGS), and the maximum number of iterations (25). Finally, in RF, we set the number of trees equal to 50, and the minimum number of samples to split a node equal to 2.

5-fold cross-validation was used throughout our tests. In particular, we used a stratified cross fold variant, which allows selecting an equal amount of ham and spam for classification, to avoid any bias towards the outcome of the classifiers (Kohavi, 1995). The performance of the learning methods is evaluated using precision, recall, F-score, Receiver Operating Characteristics (ROC) curve analysis, and Area Under the Curve (AUC). Precision (Equation 14) measures the proportion of positive results that were correctly classified. It is defined as the number of correct positive results (true positive, TP) divided by the number of all the positive results (total of TP and false positive, FP)

$$Precision = \frac{TP}{TP + FP}. \tag{14}$$

Recall is defined as the number of correct positive results (TP) divided by the number of positive results that should have been returned (total of TP and false negative, FN):

$$Recall = \frac{TP}{TP + FN}. \tag{15}$$

The F-score measures the accuracy of the test. It is a weighted average of precision and recall and it is defined as follows

$$F\text{-}score = \frac{2TP}{2TP + FP + FN}. \tag{16}$$

*ROC Curve Analysis*

ROC curves were used to measure the accuracy of the classifiers (Kyumin, 2010). Specifically, the data is represented by plotting the false positive rate (FPR) on the x-axis and the true positive rate (TPR) on the y-axis, where a steeper curve towards the y-axis is desired. The FPR (Equation 17) refers to the misclassification of the model, which should be as low as possible. It is defined as the number of misclassified ham (FP) divided by the total number of actual negatives, (i.e., FP + TN)

$$False\ Positive\ Rate = \frac{FP}{FP + TN}. \tag{17}$$

TPR is defined as the number of correctly classified spam emails (TP) divided by the total number of actual positive results, (i.e., TP + FN):

$$True\ Positive\ Rate = \frac{TP}{TP + FN} = Recall. \tag{18}$$

We used ROC curves to compare the AUC of each classifier (Hidalgo, 2006). Since the cross-validation method applied in this paper is based on five cross splits, the ROC process generated a mean calculation and a standard deviation measurement for each split, allowing the analysis of the area of variation for different partitions of the dataset.

**2.5 Development environment**

All the simulations were performed in Python 3.6. Three main modules were used to create the machine learning script used to produce the results: Natural Language Toolkit (NLTK) (Perkins, 2010), Sci-kit Learn (Pedregosa *et al*., 2011), and NumPy (Oliphant, 2006), while Matplotlib (Barrett *et al*., 2005) was used to produce the graphical output for the ROC curves. An implementation of Deep SHAP was used to compute SHAP values for machine learning models (https://github.com/slundberg/shap). The script was developed using the PyCharm integrated development environment (IDE). The experiments were run on a 6-Core i7 processor with 16GB of RAM.

# Results

In this section, we discuss the feature selection size test, the hyperparameter optimisation, and the performance of each classifier. The outcomes are compared based on precision, recall, F-score, run time required for predicting the classification outcomes, and ROC curves. Statistical tests are also reported to assess the difference between the performance of the classifiers. SHAP plots are used to investigate the features' impact on the classification outcomes.

## 3.1 Feature size selection

To investigate the impact of the number of selected features on the performance of the classifiers, we ran the models using different feature subset sizes. Figure 2(a) reports the average F-scores and average performance times recorded when running the classifiers using 10, 25, 50, 75, 100, 125, 150, and 200 features. We did not test the classifiers on a higher number of features to avoid the risk of overfitting (Roelofs *et al.,* 2019). The time performance has been scaled to show the shortest time (best performance) as 100% in the spider plot. The results show that the feature size had no impact on the average F-score of the classifiers, showing a constant value of 89%. The best time performance was achieved when using 10 and 100 features (0.0603 seconds). However, the difference between the average time recorded when using the highest number of features (200 features) and 100 features was only of the order of $10^{-3}$ seconds. Supported by the most recent research findings showing that classifiers like NB, RF, and MPNN generally perform better with supplementary features (Saeed, 2021), 200 features were selected to run the final tests.

## 3.2 Hyperparameter optimisation

The performances of MPNN and kNN were investigated based on the selection of different hyperparameters. Specifically, the number of neurons in each layer, the number of maximum iterations, and the number of hidden layers were analysed for MPNN. Three different distance algorithms were analysed for kNN, i.e., ball tree, brute force, and KD-tree.

*Multilayer Perceptron Neural Network*

The performance of MPNN was analysed by tuning three hyperparameters, including the number of neurons per hidden layer, the maximum number of iterations to perform before reporting the classification outcomes, and the number of hidden layers. Figure 2(b) reports the average F-scores and running times corresponding to five different numbers of neurons per hidden layer (25, 50, 75, 100, and 200). The plot shows that the number of neurons had an impact on the performance of the classifier in terms of time, and marginally in terms of accuracy. The best time performance was achieved using 50 neurons, where a 92% accuracy was reported in 0.067 seconds. The average F-score was equal to 91% with 25 neurons and equal to 92% with 50 neurons or more. Hence, 50 neurons per hidden layer were used for the final tests. Figure 2(c) reports the average F-scores and running times of the classifier when varying the maximum number of iterations (2000, 3500, 5000, 7500, and 10000). The results show that all the iterations acted similarly in terms of average F-score (which was constantly equal to 92%), but with 2000 iterations taking slightly more time than the other iterations (0.0084 seconds compared to 0.0083 seconds recorded with 5000 iterations or more). Considering that using 10000 iterations did have no major effect on the model's performance time, the parameter for the maximum number of iterations was set equal to 10000, allowing the network to train for longer if required.

The results reported in Figure 2(d) found no indication that the number of hidden layers affects the performance of the classifier in terms of average F-score, which was equal to 92% across the five settings (1, 2, 3, 4, and 5 hidden layers). However, based on this data, it was found that one hidden layer provided the best performance based both on F-score (92%) and time performance (0.0066 seconds).

Finally, the following hyperparameters for MPNN were selected: one hidden layer with 50 neurons, while the classifier applies a maximum of 10000 iterations before reporting the classification outcomes.

*k-Nearest Neighbour*

As detailed in the Methods section, three distance algorithms (i.e., ball tree, brute force and KD-tree) were investigated to identify the most effective parameter for the kNN classifier.
The three algorithms were compared in terms of run time, precision, recall, and F-score. The experiments were repeated 30 times using different partitions of the dataset based on a 5-fold cross-validation approach. Figure 3 shows that brute force was the best performing algorithm in terms of precision, recall, and F-score, with median values equal to 86.21%, 84.94%, and 86.46%, respectively. When using brute force as a distance algorithm, kNN also showed the lowest performance time (median time equal to 0.2118 seconds). Hence, brute force was used as a distance algorithm for the kNN classifier for the final tests.

## 3.3 Classifiers performances and ROC curves

We compared the performance of the 12 classifiers in terms of prediction run time, precision, recall, F-score (Table 3), and AUC-ROC curves (Figure 4). The optimal performance point in a ROC curve diagram is the top-left corner, while the AUC provides a summary of the performance of the method.
Table 3 summarises the results for all the classifiers reporting the values of the comparison metrics. RF and XGBoost slightly outperformed MPNN in terms of F-score (94%, 94%, and 92% respectively). However, MPNN showed the shortest run time (0.007 seconds) compared to XGBoost (0.017 seconds) and RF (0.025 seconds). By comparing the AUC of the three models (Figure 4), XGBoost reported an average AUC higher than RF (0.98±0.01 and 0.97±0.01, respectively), suggesting that XGBoost might be preferable over RF for larger datasets. MPNN and Bernoulli NB showed the third-highest mean AUC (0.96±0.02, Figure 4).

LR resulted to be the fastest classifier (0.006 seconds) with precision, recall and F-score equal to 89%, 87%, and 89%, respectively (Table 3), and AUC equal to 0.95±0.02 (Figure 4). The same F-score and AUC were recorded with linear SVM; however, this classifier reported a significantly longer prediction time (0.212 seconds).
Bernoulli NB outperformed the other two NB algorithms with a performance time of 0.008 seconds and precision, recall and F-score equal to 91%. The F-score of Gaussian NB was 87%, with 87% precision and 85% recall. Multinomial NB reported the lowest F-score among the three NB methods (85%), with 85% precision, and 85% recall. In terms of AUC, Figure 4 shows that Bernoulli NB had also a higher AUC (0.96±0.02) compared to Gaussian NB (0.91±0.03), and Multinomial NB (0.93±0.03).

Linear SVM was the best performing SVM classifier with an F-score of 89%, followed by RBF SVM (71%), sigmoidal SVM (70%), and polynomial SVM (67%). In terms of performance time, linear SVM was also the fastest algorithm of the SVM classifiers with a performance time of 0.212 seconds. The ROC curves of the SVM classifiers in Figure 4 support these results showing the linear SVM as the classifier with the highest mean AUC (0.95±0.02), followed by sigmoidal SVM and RBF SVM (both with AUC equal to 0.88±0.03). In terms of AUC, polynomial SVM was the worst-performing classifier with AUC equal to 0.76±0.14.

## 3.4 Comparative analysis

To provide a robust analysis of the performance of the 12 machine learning models and interpret the classification outcomes, statistical and explainability tests were performed.

The sections below report the details and results of the test performed to statistically validate and interpret our results.

### 3.4.1 Statistical significance

To determine the statistical significance of the F-scores obtained by the 12 machine learning models, paired and pair-wise t-tests were applied. Specifically, adjusted p-values using Bonferroni correction were used to assess the statistical difference at a 5% significance level. The tests were repeated 20 times and the F-scores of each method were compared.

A visual representation of the t-test results is presented in Figure 5(a). The boxplots of the F-scores for the top-3 best performing models (i.e., RF, XGBoost, and MPNN) are reported. The statistical difference is indicated by the horizontal bars, where **** shows that the adjusted p-values resulting from the t-test were less than 0.0001. The plot shows that there is a statistically significant difference between MPNN and the two other models, while there is no significant difference between RF and XGBoost. However, the results in Table 3 show that XGBoost achieves the same performance in a shorter time (0.017 seconds compared to 0.025 seconds) suggesting that this method might be preferable over RF for larger datasets. The results of the pairwise comparisons across all the 12 methods show that there is a significant difference across most of the models except 5 pairs, including RF and XGBoost, Gaussian NB and kNN, Gaussian NB and Multinomial NB, kNN and Multinomial NB, and LR and linear SVM. The details of the t-tests statistic are reported in Supplementary File 1.

### 3.4.2 Model interpretability

To quantify the feature contributions on the classification outcomes, and provide an interpretation of the results, we used the SHAP (SHapley Additive exPlanations) method (Lundberg *et al.,* 2017). Specifically, we used an implementation of Deep SHAP, an algorithm to compute SHAP values for machine learning models https://github.com/slundberg/shap).

Figures 5(b)-(d) show the SHAP value plots used to investigate the impact of the features on the models' outcomes and provide recommendations on the selection of the optimal features/models. The SHAP plots for the top-three best performing models are reported, i.e., (b) RF, (c) XGBoost, and (d) MPNN. Each point in the plot represents an instance of the test set. Variables are ranked in descending order and the top-10

features for each model are reported. The horizontal location shows whether the effect of that instance is associated with a spam/not spam prediction. The colours show whether the occurrence of that feature is high (red) or low (blue) for each instance.

Figure 5(b) presents the SHAP plot related to RF. The results show that this model selects words that are commonly found in all emails and a low occurrence of most of them has no impact on the model's output (blue points corresponding to a zero SHAP value). These include words like *gas*, *thanks*, and *schedule*. However, a high frequency of these words has a negative impact on the classification outcomes, i.e., a high occurrence of these words negatively correlates with the classification of the instance as spam. On the contrary, a high occurrence of the words *http* and *price* has a positive impact on the classification of the instance as spam. It is worth noting that *http* refers to the unsecured protocol for websites, which is commonly used due to insufficient signature applied to a scam website. In fact, when the dataset was released in 2007, most websites were commonly *http* only, as security was not given the same priority that receives today.

Figure 5(c) shows the SHAP plot of the XGBoost classifier. A high occurrence of the top-9 features had a positive impact on the classification of the instances as spam. Specifically, 4 high-frequency features that positively correlate with the classification outcomes were identified, i.e., *future*, *order*, *software*, and *phone*. The word *thanks* was the only feature contributing towards the classification of the instance as non-spam, showing that XGBoost relies more on spam-related features than other models.

Figure 5(d) reports the SHAP plot of the MPNN classifier. The model selects among the top features words that are related to the company's operation. Some of these words, such as *gas*, could be considered noise or business-specific due to Enron corporation being an energy company. This justifies the low impact on the model's output of the low occurring features (blue points in proximity of zero SHAP value). However, the high frequency of the same features shows a higher correlation (either positive or negative) with the classification outcomes. In particular, the plot shows that the high frequency of the words *best* and *account* has a positive impact on the classification outcome (i.e., there is a positive correlation between the high occurrence of these words and the classification of the email as spam).

By analysing the features selected across the three models, it is worth noting that the word *please* has been identified among the top-3 features by all the models. Specifically, all the SHAP plots show that a low frequency of this word has a positive impact on the classification outcome (there is a weak positive correlation between the low occurrence of the word *please* and the classification of the instance as spam). A high occurrence of the word *thanks* has been classified as having a negative impact on the non-spam classification outcomes by RF and XGBoost (Figures 5(b) and (c), respectively). This shows that a high occurrence of this word correlates negatively with the classification of the email as spam. The SHAP plots of the remaining classifiers showed a similar trend as reported in Supplementary Figures 1 and 2.

Overall, Random Forest and MPNN seem to rely on complex words for the classification of each instance, while XGBoost identifies features/words that are more commonly used in spam emails.

### 3.4.3 Impact of the preprocessing steps

In the proposed approach, we included lemmatisation as a preprocessing step. Lemmatisation is the process of grouping together the inflected forms of a word so that they can be analysed as a single item reducing the complexity of the dataset. To test the effectiveness of this step, we repeated the full pipeline without applying any preprocessing step to the original dataset. Figure 5(e) reports the results showing a clear improvement of the performance in terms of F-scores across the top-6 best performing models. The figure shows that when including the preprocessing steps, the 6 models achieved more than a two-fold performance improvement, supporting the effectiveness of the proposed preprocessing steps. Similar results were shown by the other models as reported in Supplementary File 2.

## Conclusion

We proposed and tested a pipeline to compare and explain the classification outcomes of 12 machine learning models. We applied the pipeline for optimising and testing the models in a spam filtering context, with lemmatisation and noise-reduction techniques as preprocessing steps. The pipeline, which we make publicly available, was developed to compare the performance of the classifiers in terms of precision, recall, F-score, and ROC curves.

Specifically, we used the Enron spam corpus with the machine learning algorithms to achieve a reliable spam classification, reporting an F-score of 94%. Along with this outcome, the importance of hyperparameters was highlighted and resulted to have a major impact on MPNN by reducing the time requirement and increasing the accuracy. The results found that XGBoost was the best performing classifier, displaying a 94% F-score, while only requiring 0.017 seconds. RF presented similar results in terms of F-score (94%); however, the classifier required 0.025 seconds to generate the classification outcomes. MPNN was reported to be the third-best classifier, presenting an average F-score of 92%, while recording the second-lowest time performance of 0.007 seconds. The statistical analysis and interpretability investigation showed that even when there was not a statistically significant difference between the models' performances, analysing the features' impact can provide valuable insights on the different classification approaches.

While the pipeline was effective in explaining the models' results, there are some points for discussion. Firstly, the performance results of the polynomial SVM algorithm reports this to be the worst-performing classifier (Table 3), which suggests that a more thorough parameter optimisation and a less strong preprocessing are needed to improve the performance on this dataset. Secondly, the specific dataset used for our pipeline might positively or negatively affect the performance of the classifiers. This result could be due to the following reasons: (i) the Enron corpus is known to be an enterprise spam corpus, thus resulting in the possibility of tailored spam towards the now late Enron company, and (ii) bias towards the dataset, specifically regarding the data of all emails and placing certain words that allow the machine learning classifiers to easily detect them. Overall, our results support the importance of choosing bespoke algorithms for different problem domains.

In our pipeline, lemmatisation and noise reduction techniques were applied with hyperparameter optimisation, reporting better performance metrics. The aim of noise reduction in the specific order was to identify how this process can improve the performance of the text classification. Specifically, the proposed preprocessing techniques allowed more than a two-fold performance improvement across most of the 12 machine learning classifiers. The same pipeline appears therefore as a promising tool to be applied to other text-classification problems, where a set of NLP-based and noise-reduction preprocessing can improve the classifiers' performance. Future directions include the evaluation of other machine learning algorithms that apply supervised learning along with investing time into optimising the hyperparameters of these algorithms. The issues discussed above, related to the possibility of bias with public datasets, should also be addressed in future works.

## Acknowledgements

AO would like to thank the support from the Earlier.org Breast Cancer Award. CA would like to acknowledge the support of UKRI Research England's THYME project, and a Children's Liver Disease Foundation Research Grant.

| Set Name | Total Emails Ham | Total Emails Spam | Theoretical ratio Ham/Spam | Actual ratio Ham/Spam | Subset Full total |
|---|---|---|---|---|---|
| Enron 1 | 3672 | 1500 | 3:1 | 2.45 | 5172 |
| Enron 2 | 4361 | 1496 | 3:1 | 2.91 | 5,857 |
| Enron 3 | 4012 | 1500 | 3:1 | 2.67 | 5512 |
| Enron 4 | 1500 | 4500 | 1:3 | 0.33 | 6000 |
| Enron 5 | 1500 | 3675 | 1:3 | 0.41 | 5175 |
| Enron 6 | 1500 | 4500 | 1:3 | 0.33 | 6000 |
| Full Set | 16545 | 17171 | - | 0.96 | 33716 |
| Applied Set | 16545 | 16545 | 1:1 | 1 | 33090 |

**Table 1. Enron dataset information.** The table shows the number of ham (non-spam) and spam emails for each subset of the Enron corpus dataset. For this study, we used the applied set where all the 16545 ham emails available in the full set were considered, and the same number of spam emails was selected. Specifically, we used a one-time randomisation selection on spam emails to achieve an actual ratio of one, which was used throughout the course of the training and testing phases.

| Algorithms | Parameters | Values |
|---|---|---|
| Support Vector Machine | kernel | **linear, polynomial, sigmoidal, radial basis function** |
| | max iterations | 5 |
| | degree | 3 |
| | *c* - penalty parameter | 1 |
| k- Nearest Neighbour | algorithm | **brute force**, ball tree, KD-tree |
| | leaf size | 10 |
| | *p* – power parameter | 1 |
| | number of neighbours | 5 |
| Multi-layer perceptron neural network | number of neurons | 25, **50**, 75, 100, 200 |
| | hidden layers | **1**, 2, 3, 4, 5 |
| | solver | LBFGS |
| | max iterations | 2000, 3500, 5000, 7500, **10000** |
| Logistic regression | solver | LBFGS |
| | max iterations | 25 |
| Random forest | number of trees | 50 |
| | min number of samples to split an internal node | 2 |

**Table 2. List of hyperparameters.** For each classifier, we set different hyperparameters based on the analysis reported in the Methods section. The table shows the values used for running the final tests. Where multiple values are provided, the hyperparameters selected for running the final tests are shown in bold. For SVM, all the kernels were selected for the final comparisons. The parameters not listed in the table have been set equal to the default values of the specific classifier.

| Classifier | Time (s) | Precision | Recall | F-score |
|---|---|---|---|---|
| kNN | 0.217 | 82% | 81% | 83% |
| MPNN | 0.007 | 92% | 91% | 92% |
| Logistic Regression | **0.006** | 89% | 87% | 89% |
| Random Forest | 0.025 | **94%** | **94%** | **94%** |
| XGBoost | 0.017 | **94%** | 93% | **94%** |
| Multinomial NB | 0.009 | 85% | 85% | 85% |
| Gaussian NB | 0.011 | 87% | 85% | 87% |
| Bernoulli NB | 0.008 | 91% | 91% | 91% |
| RBF SVM | 3.661 | 77% | 58% | 71% |
| Linear SVM | 0.212 | 89% | 88% | 89% |
| Poly SVM | 0.675 | 25% | 50% | 67% |
| Sigmoid SVM | 0.806 | 76% | 56% | 70% |

**Table 3. Performance of the 12 machine learning models.** The table reports the performance results for the 12 algorithms used in our investigation in terms of prediction time, precision, recall, and F-score. For each algorithm, the time to classify the instances in the test set is reported in seconds. The values in bold show the best performance for each metric.
Random forest and XGBoost achieved the highest performance in terms of F-score. However, the performance time of random forest was higher than XGBoost (0.025 seconds compared to 0.017 seconds). Overall, the optimal time performance was obtained by MPNN, with an F-score of 92% in 0.007 seconds.

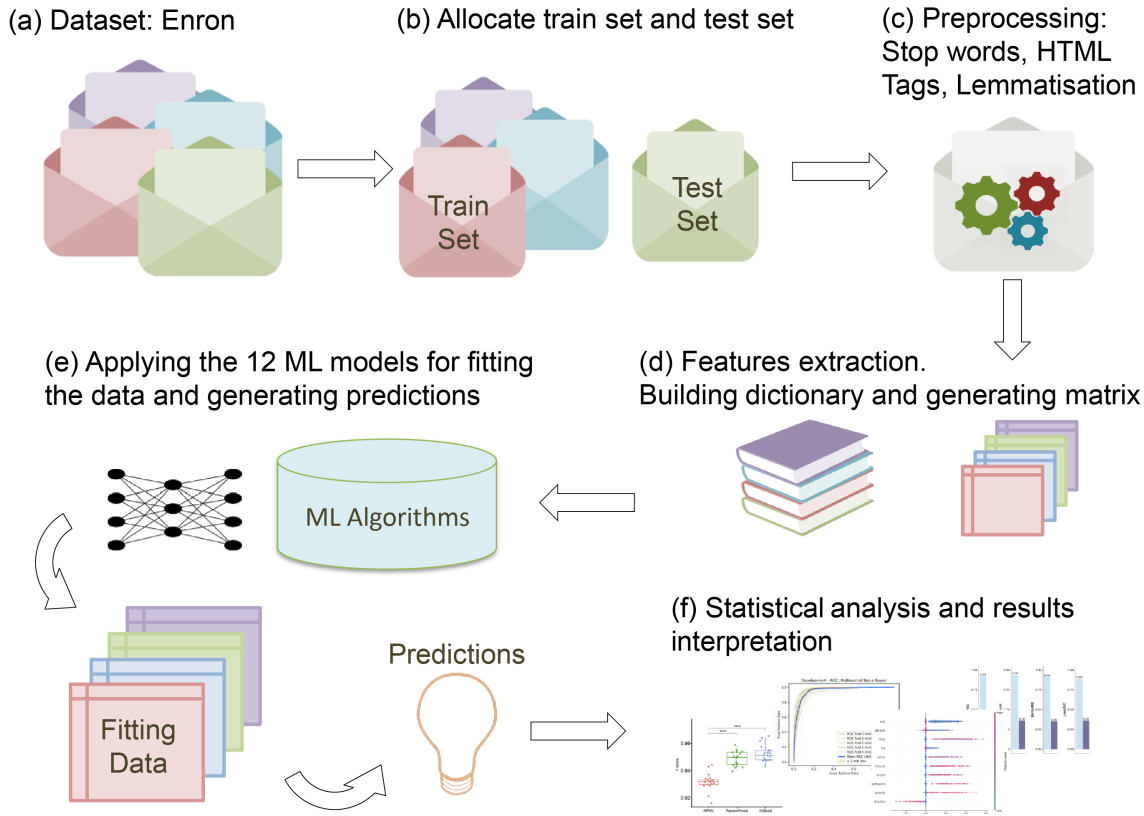

**Figure 1. Pipeline diagram of the preprocessing and classification steps.** The pipeline shows the process applied for investigating the performance of 12 classifiers. (a) The Enron dataset is selected for the study; (b) the train and test sets are allocated (70% train set and 30% test set); (c) the preprocessing stage is then applied, including removal of stop words, HTML tag and lemmatisation; (d) the features are then extracted for generating a dictionary, based on the number of most occurring features; (e) finally, the 12 algorithms receive the matrices for fitting the data and predicting the classification outcomes; (f) comparative analysis is then performed to assess the statistical significance of the results and provide an accurate interpretation of the machine learning classification outcomes.

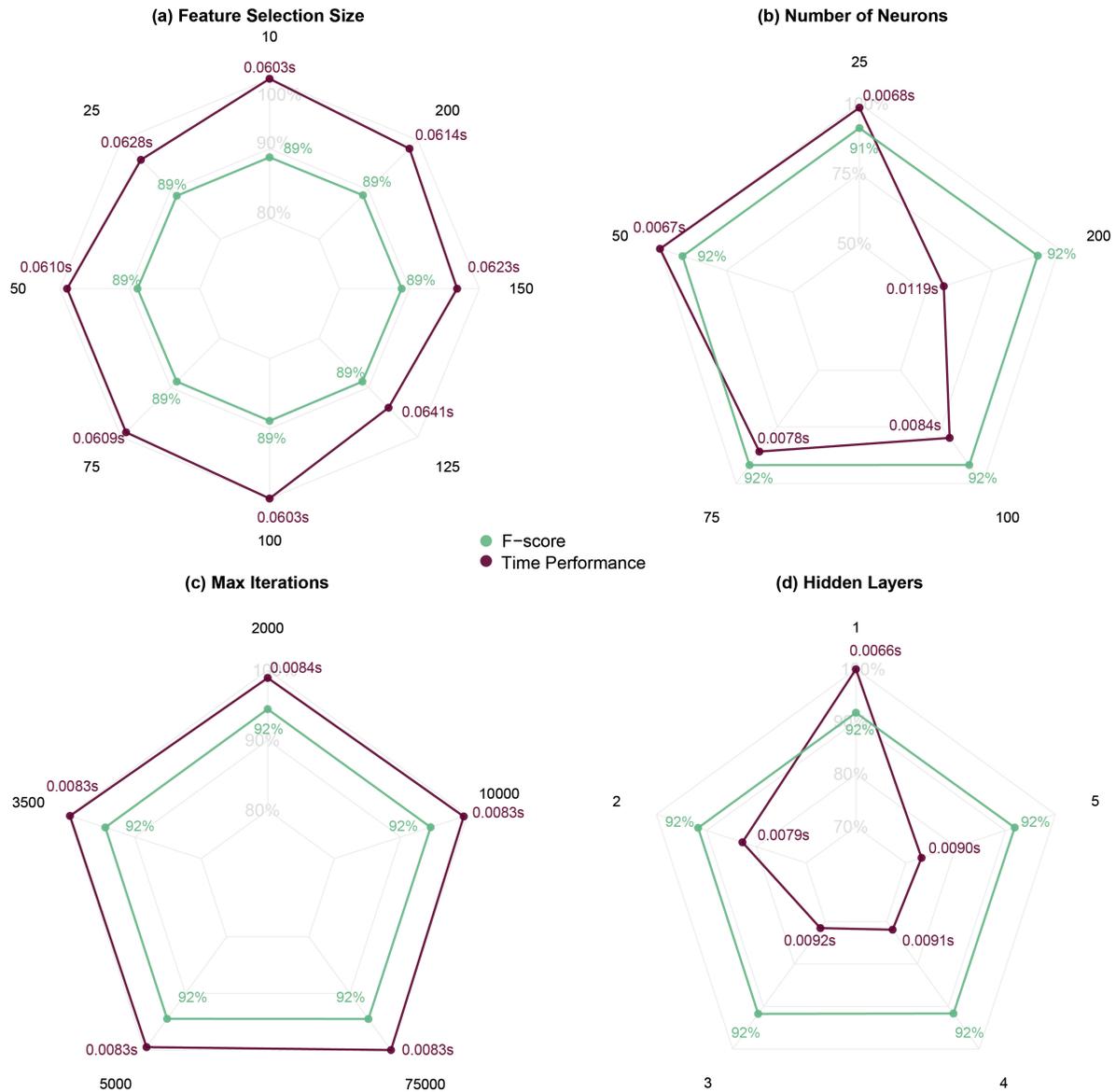

**Figure 2. Feature size selection and hyperparameter optimisation.** (a) Feature selection test. The plot reports the average F-score and time required for prediction for different subset sizes of the features set (10, 25, 50, 75, 100, 125, 150, and 200). The time has been scaled to show 100% as the optimal performance (corresponding to the shortest time). The number of features had no impact on the average F-score, which was equal to 89% across the eight settings. The best time performance was recorded with 10 and 100 features (0.0603 seconds). However, considering that algorithms like kNN and MPNN require larger datasets for better performance and that the difference between 200 features and 100 features was only of the order of $10^{-3}$ seconds, the feature set size selected for the final tests was 200 features. (b), (c) and (d) report the performance of MPNN when tuning different hyperparameters, i.e., number of neurons (b), maximum number of iterations (c), and number of hidden layers (d). (b) The best performance was obtained with 50 neurons (92% F-score and 0.067 seconds run time), which was the number of neurons used for running the final test. (c) Altering the number of iterations had minimal impact on F-score and time performance. Hence, 10000 iterations were selected

for the final test, allowing the network to be trained for longer if required. (d) When tuning the number of layers, the F-score performance remained constantly equal to 92%. However, when using one hidden layer the algorithm showed the best performance in terms of time required. For this reason, one hidden layer was used for running the final tests.

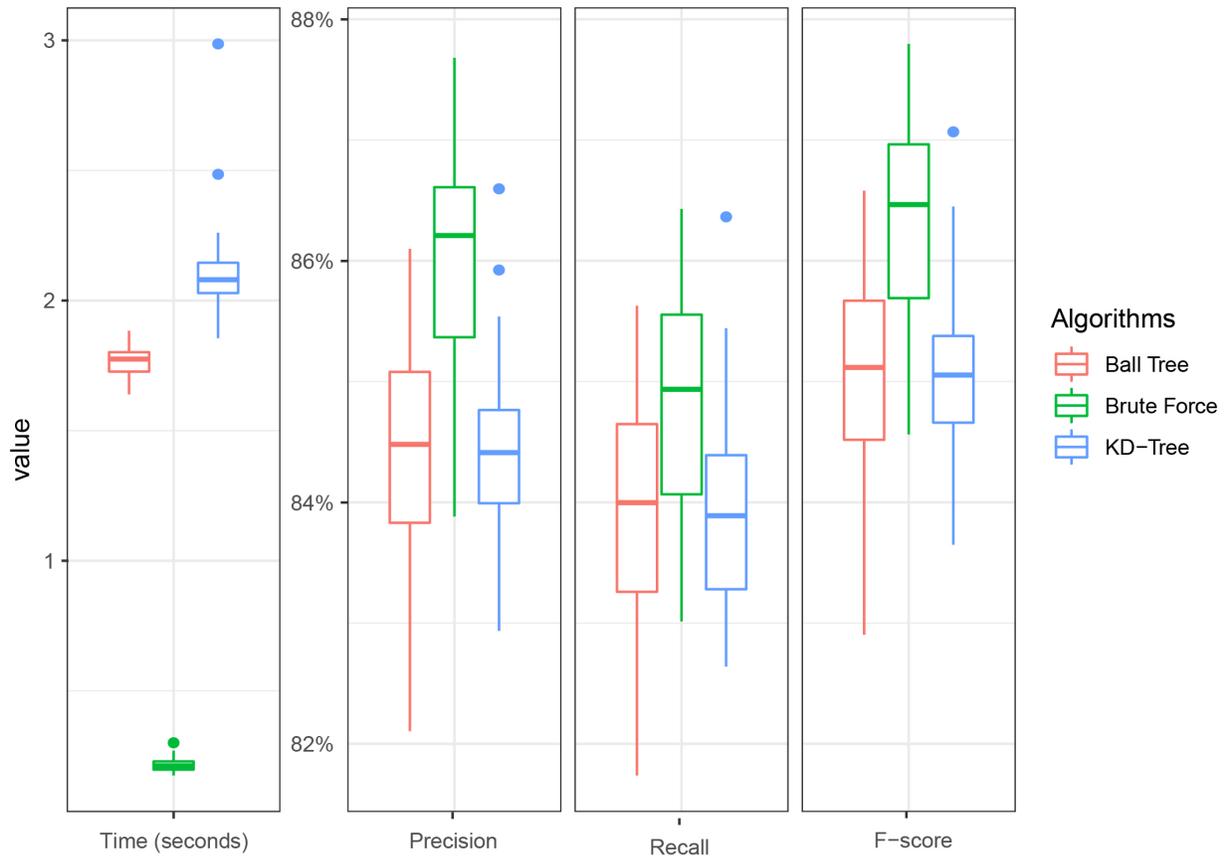

**Figure 3. kNN classifier performance using three distance algorithms.** Three kNN algorithms (ball tree, brute force, and KD-tree) were compared in terms of performing time, precision, recall, and F-score. The experiments were repeated 30 times using different partitions of the dataset based on a 5-fold cross-validation approach. Brute force was the best-performing algorithms in terms of precision, recall and F-score (median values were equal to 86.21%, 84.94%, and 86.46%, respectively). This algorithm also showed the optimal performing time (median value equal to 0.2128 seconds) and it was selected for the final tests.

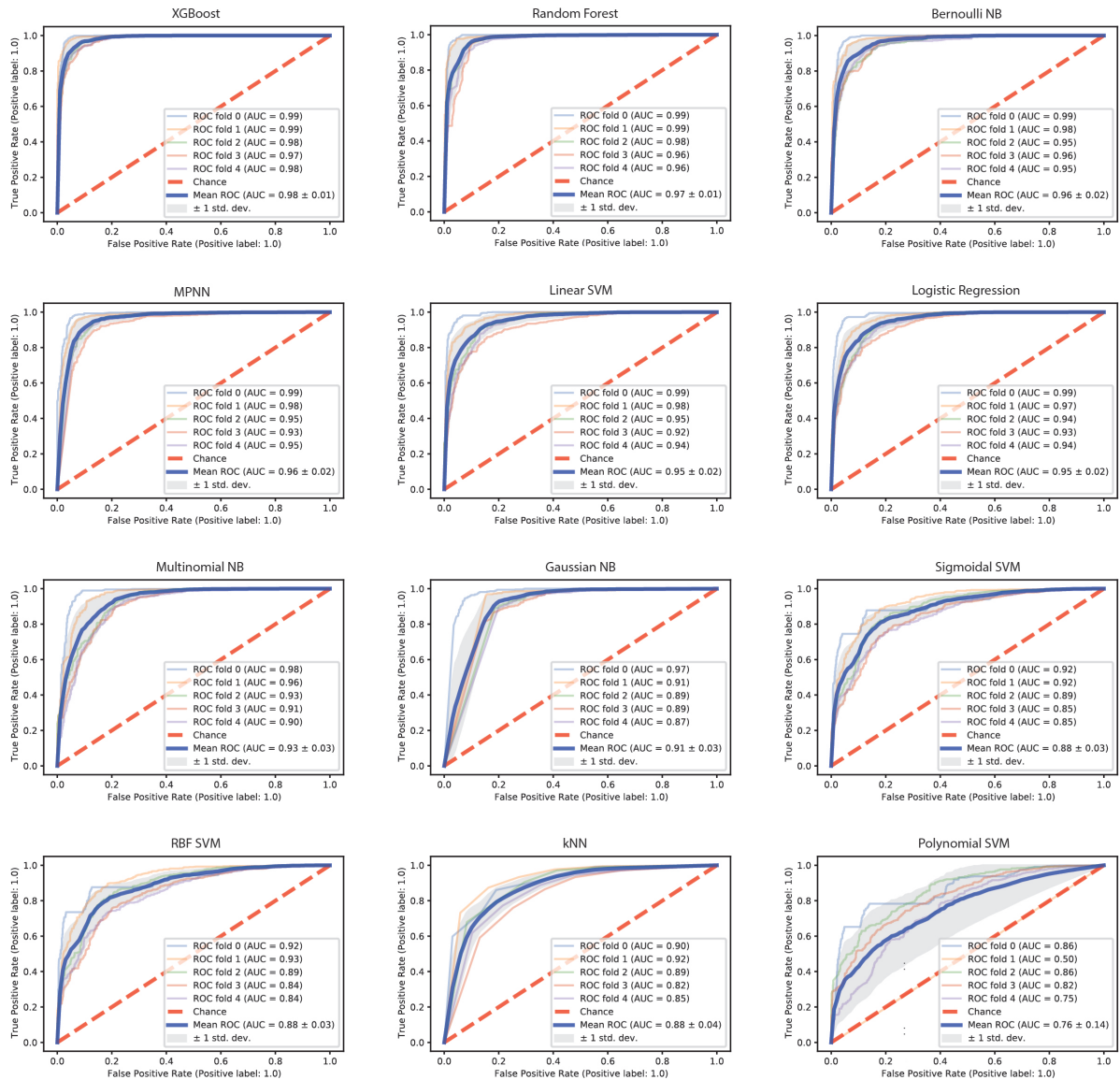

**Figure 4. ROC Curves of the 12 classifiers in order of mean Area Under the Curve (AUC).** From the analysis of the ROC curves, XGBoost was the best performing algorithm with a mean AUC equal to 0.98±0.01, followed by random forest (0.97±0.01), Bernoulli NB (0.96±0.02), and MPNN (0.96±0.02). The top-performing SVM model was Linear SVM with AUC equal to 0.95±0.02. The other three SVM classifiers were among the bottom-4 models with the kNN classifier showing a mean AUC of 0.88±0.03 or lower. Even if logistic regression was the top-performing algorithm in terms of run time (see Table 3), its ROC curve shows that its classification performance was worst than the top-performing model (AUC equal to 0.95±0.02). Among the three Naïve Bayes classifiers, Bernoulli NB was the top-performing model, followed by Multinomial NB and Gaussian NB (AUC equal to 0.93±0.03 and 0.91±0.02, respectively).

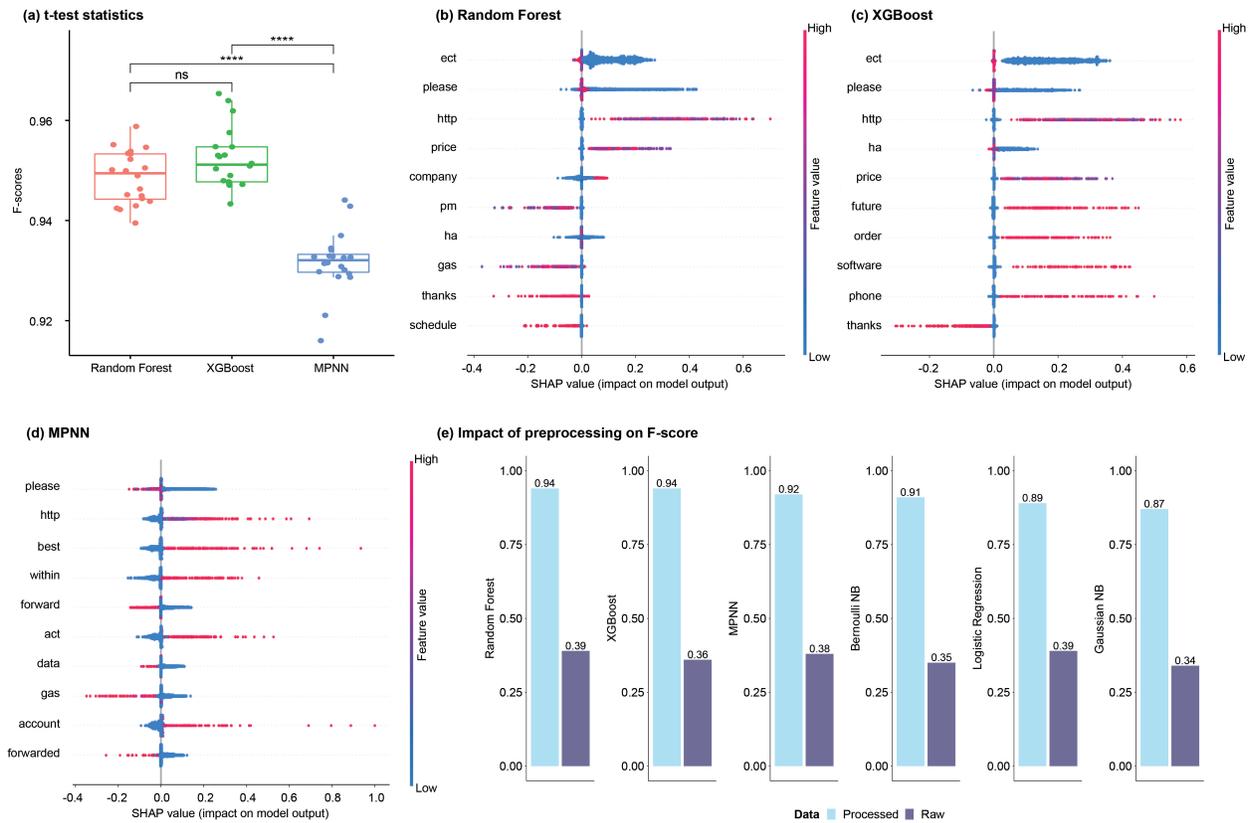

**Figure 5. Statistical analysis and results interpretation.** (a) Results of the paired and pair-wise t-test statistics performed to assess the statistical difference between the models' performance in terms of F-score (at a 5% statistical level). Bonferroni correction has been used to compute the final p-values. The figure reports the test results of the top-3 best performing models. The performance scores of random forest and XGBoost were statistically different from MPNN's results. **** shows that the adjusted p-values resulting from the t-test were lower than 0.0001. The results related to the remaining models, reported in Supplementary File 1, show that only 5 comparisons out of 66 were not significant. (b)-(d) SHAP plots showing the features' contribution to the models' outcome for the top-3 best performing models. The word *please* has been identified among the top-3 features by all three models. Specifically, the low occurrence of this word has a positive impact on the classification outcome (there is a weak positive correlation between the low occurrence of the word *please* and the classification of the email as spam). Overall, Random Forest and MPNN seem to rely on complex words for the classification of each instance, while XGBoost identifies features/words that are more commonly used in spam emails. (f) Impact of preprocessing steps on the performance of the top-6 best-performing classifiers. All 6 models achieved more than a 2-fold improvement in terms of F-score when applying the preprocessing steps proposed in our work. Similar results were shown by the other models (more details are reported in Supplementary File 2).

# Supplementary Material

### Supplementary File 1. T-test Results for all Models
This file contains the statistical details of the paired and pair-wise t-test statistics.

### Supplementary File 2. Impact of preprocessing steps on the performance of the classifiers.
This file contains the F-scores obtained when running the 12 machine learning models on the raw data (Performance with raw data tab), and when applying the preprocessing steps in the initial phase of the pipeline (Performance with preprocessing tab).

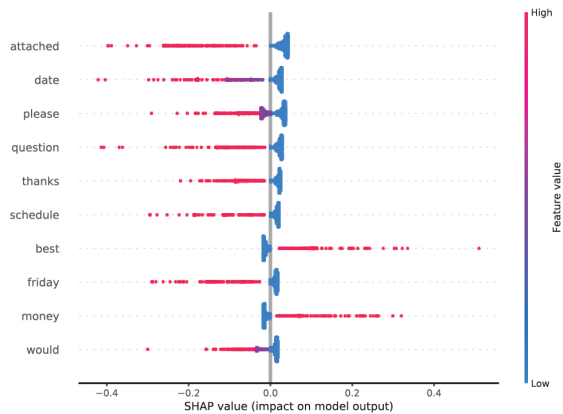
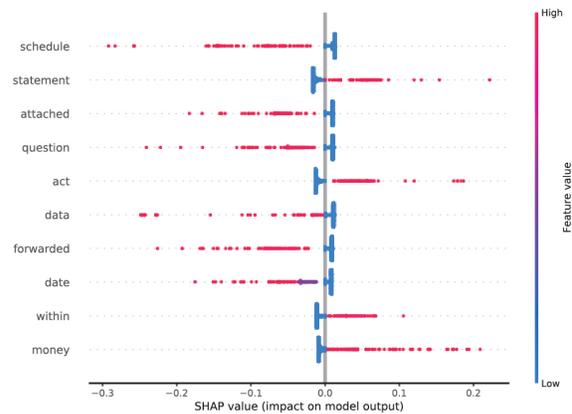
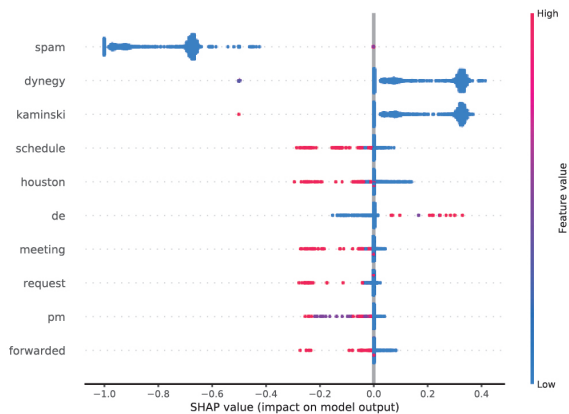
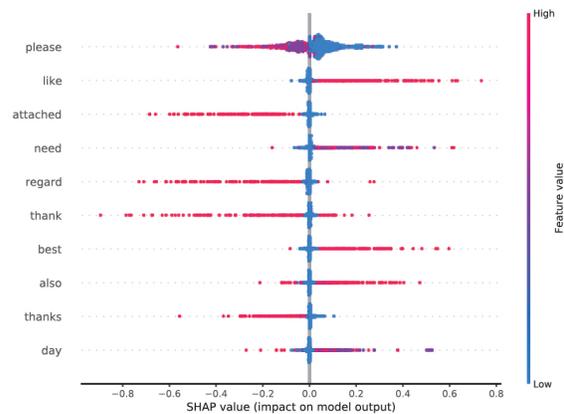

**Supplementary Figure 1. SHAP plots of (a) Logistic Regression, (b) Multinomial NB, (c) Gaussian NB, and (d) kNN.** (a) SHAP plot associated with Logistic Regression, which identifies words that occur in both non-spam and spam. This justifies the low impact on the model's output of the low occurrence of these features (blue points in proximity of a zero SHAP value). However, a high occurrence of the same features has a higher correlation (either positive or negative) with the classification outcomes. In particular, the plot shows that a high frequency of

the words *best* and *money* has a positive impact on the classification outcome (i.e., there is a positive correlation between the high frequency of these words and the classification of the email as spam). Similar behaviour is shown in the SHAP plots associated with Multinomial NB (b) where a low frequency of the selected features has almost no impact on the model's output (blue points in proximity of zero SHAP value). However, a high frequency of words like *statement, act,* and *money* has a high impact on the classification of the instance as spam. Different features have been selected by Gaussian NB (c), where the word *spam* has been identified as the top feature. Specifically, low occurrences of this word have a high impact on the classification of the instance as non-spam. (d) SHAP plot related to the kNN classifier. The results show that this model selects words that are commonly found in all emails and low values of the selected features have no impact on the model outcomes, as observed for logistic regression. A high occurrence of the words *like* and *best* has also been identified as having a positive contribution to the classification of the instance as spam. Words like *attached, regard, thank,* and *thanks,* had a negative impact on the classification of the instances as spam.

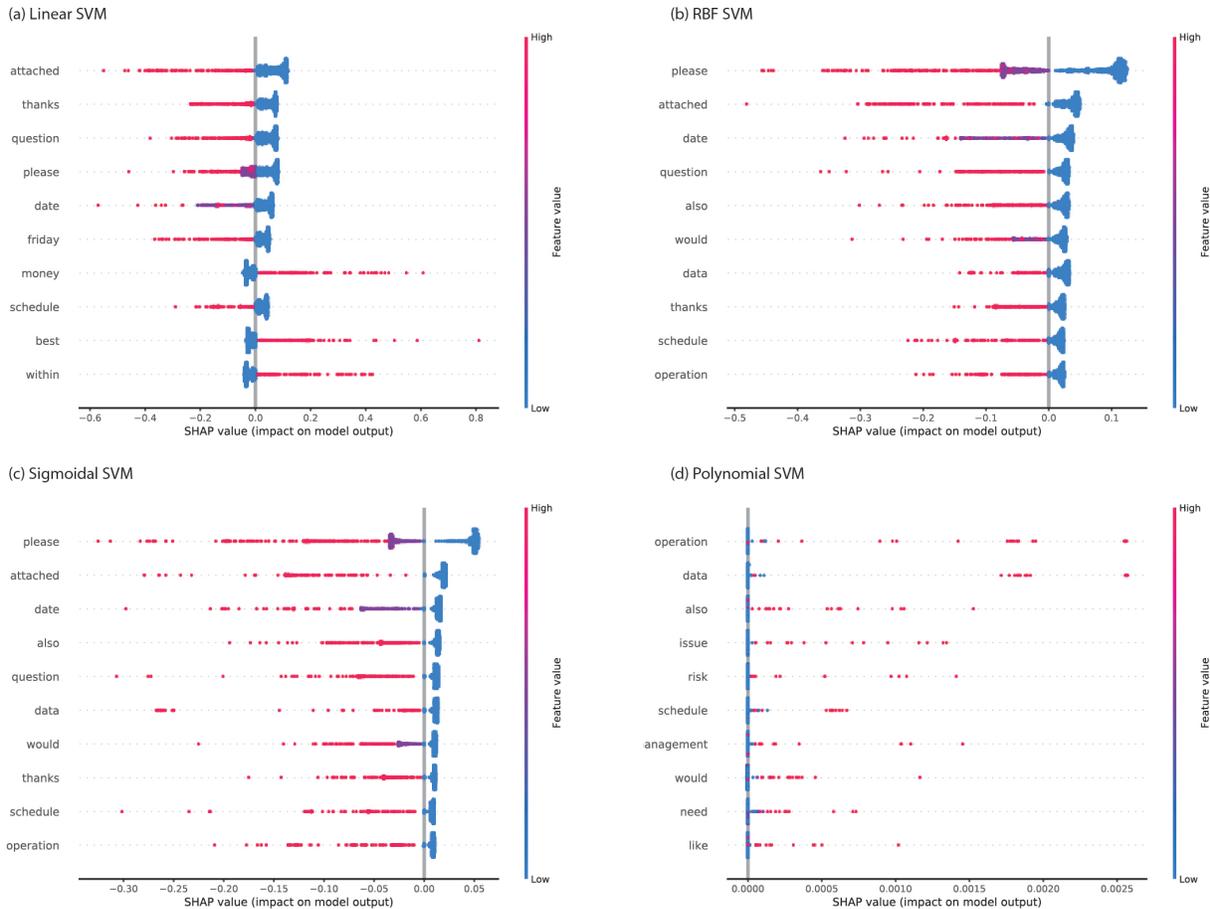

**Supplementary Figure 2. SHAP plots of the SVM classifiers. (a) Linear SVM, (b) RBF SVM, (c) Sigmoidal SVM, and (d) Polynomial SVM.** The plots show that linear SVM (a), RBF SVM (b), and sigmoidal SVM (c) identified most features that have a negative impact on the classification of the instances as spam. These include words like *please, attached,* and *thanks,* whose low occurrence has almost no impact on the classification outcomes (blue values near to zero SHAP value). (a) Linear SVM also identified words like *money, best,* and *within,* which positively contributed to the classification of the input as spam. (c)-(d) RBF SVM and sigmoidal SVM selected the same top-10 features, whose high occurrence correlated positively with the classification of the instances as spam. (d) Polynomial SVM selected different features compared to other SVM models. All the selected features had a positive impact on the model's classification outcomes, even if the contribution of these features was much smaller compared to the other models (all the SHAP values are less than 0.0025). This is due to the low precision achieved by this model (25%) as shown in Table 3 in the main text.
Overall, these results suggest that (b) RBF SVM and (c) sigmoidal SVM seem to rely on words that are more commonly used in non-spam emails, while (a) linear SVM identifies features that can be found both in ham and spam.